\documentclass[ aps,%
floatfix,%
final,%
notitlepage,%
oneside,%
onecolumn,%
nobibnotes,%
nofootinbib,%
superscriptaddress,%
showpacs,%
]%
{revtex4}

\usepackage{epsfig}
\usepackage{graphics}

\newcommand{\epem}{e^+e^-}

\newcommand{\beq}{\begin{eqnarray}}\newcommand{\eeq}{\end{eqnarray}}
\newcommand{\beqa}{\begin{eqnarray*}}\newcommand{\eeqa}{\end{eqnarray*}}

\begin{document}

\title{The process $\epem\to J/ \psi X(3940)$  at $\sqrt s = 10.6$ GeV in the framework of light cone formalism.}
\author{V.V. Braguta}
\email{braguta@mail.ru}

\author{A.K. Likhoded}
\email{Likhoded@ihep.ru}

\author{A.V. Luchinsky}
\email{Alexey.Luchinsky@ihep.ru}
\affiliation{Institute for High Energy Physics, Protvino, Russia}

\begin{abstract}
This paper is devoted to the study of the process $\epem\to J/ \psi X(3940)$ in the framework of light cone formalism. 
In our calculation two hypotheses about the structure
of $X(3940)$ meson are considered: $X(3940)$ is $3^1$S$_0$ state and $X(3940)$ is one of $2^3$P states. The former 
hypothesis leads to a good agreements with the cross section measured at the experiment. 
As to the latter one, it is proposed a mechanism that allows one to understand 
the suppression of 2P mesons production in hard processes.
\end{abstract}

\pacs{
12.38.-t,  % Quantum chromodynamics ... ... Quarks, gluons, and QCD in nuclei and nuclear
% processes
12.38.Bx,  % Perturbative calculations
13.66.Bc,  % Hadron production in e-e+ interactions
13.25.Gv % Decays of J/psi, Upsilon, and other quarkonia
}

\maketitle

\newcommand{\ins}[1]{\underline{#1}}
\newcommand{\subs}[2]{\underline{#2}}
%%%%%%%%%%%%%%%%%%%%%%%%%%%%%%%%%%%%%%%%%%%%%%%
\section{Introduction}

There are a number of charmonium-like mesons discovered recently in different experiments (see review \cite{Swanson:2006st}). 
Our paper is devoted to a charmonium-like meson $X(3940)$ 
discovered in inclusive process $e^+e^- \to J/ \psi+$anything at Belle \cite{Abe:2005hd}. The distribution of masses recoiling 
against the reconstructed $J/ \psi$ in $e^+e^- \to J/ \psi+$anything events measured at this experiment is
shown in Fig. 1. 

\begin{figure}[h]
\begin{picture}(150, 50)
\put(-50,-100){\epsfxsize=9cm \epsfbox{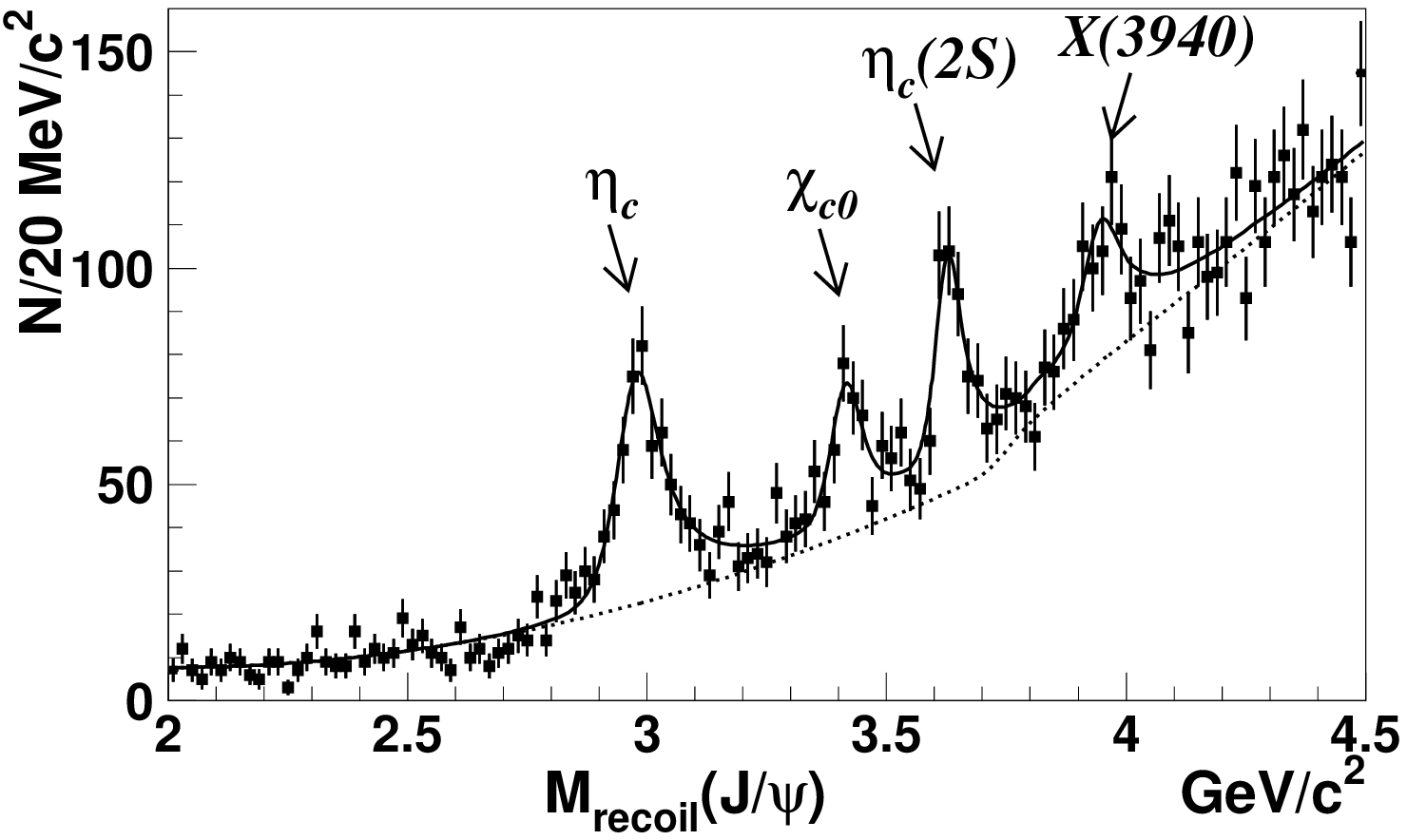}}
\put(-170, -110){\bf{Fig. 1: The distribution of masses recoiling against the reconstructed $J/ \psi$ in $e^+e^- \to J/ \psi+$anything events.}}
\end{picture}
\end{figure}

\vspace*{4cm}

Evidently the theory that pretends to the understanding of the mechanism of $X(3940)$ meson production in $\epem$-annihilation must also describe 
well the processes $e^+e^- \to J/ \psi \eta_c, J/ \psi \eta_c', J/ \psi \chi_{c0}$ measured at Belle. 
Considering charmonium mesons as a nonrelativistic bound states of $c \bar c$ pair one tried to apply 
NRQCD to these processes \cite{Braaten:2002fi}, but the results of these calculation proved to be in contradiction with 
Belle and BaBar measurements \cite{Abe:2004ww, Aubert:2005tj}.

Another approach to the prediction of the cross sections of these processes is light cone formalism. 
The authors of paper \cite{Bondar:2004sv} calculated the process $e^+e^- \to J/ \psi \eta_c$ in the framework of this 
approach. Despite of the uncertainties the agreement with the experiments was rather good. 
Further progress in understanding of exclusive double charmonium production in $e^+e^-$ annihilation was connected with 
papers \cite{Braguta:2005kr} and \cite{Braguta:2006nf}. In the former paper the process $e^+e^- \to J/ \psi \eta_c'$ was calculated.
The latter one predicts the cross section of the process $e^+e^- \to J/ \psi \chi_{c0}$. The values of the 
cross sections obtained in these papers are in good agreement with experimental results. In addition to good agreement 
with the experiments the application of light cone formalism allows one to understand 
that the wave functions of charmonium mesons are too wide to describe double charmonium production in the framework of NRQCD. 

In our paper we will consider $X(3940)$ meson in the framework of conventional quark model and suppose that this meson is
an excitation of charmonium meson already seen at Belle experiment. 
There are tree mesons $\eta_c(1^1S_0), \eta_c'(2^1S_0), \chi_{c0}(1^3P_0)$ in Fig. 1. 
Actually the peak labeled by $\chi_{c0}(1^3P_0)$ can arise from $\chi_{c0}(1^3P_0), \chi_{c1}(1^3P_1), \chi_{c2}(1^3P_2)$ mesons, 
although $\chi_{c0}$ gives dominant contribution. 
So two scenarios must be considered: $X(3940)$ 
is $\eta_c''(3^1S_0)$ state and $X(3940)$ is one of $\chi'_{c0}(2^3P_0), \chi'_{c1}(2^3P_1), \chi'_{c2}(2^3P_2)$ states.

This paper is organized as follows. In section 2 the model for the light cone wave functions used in our calculation
is considered. 
Section 3 is devoted to the calculation of the cross section of $X(3940)$ meson production. 
In last section we discuss results obtained in our paper.

\section{The light cone wave functions of $^1S_0, ^3S_1$ and $^3P_0$ the mesons.}
In this section the light cone wave functions of $^1S_0, ^3S_1$ and $^3P_0$ mesons will be considered. 
The functions used in our calculation are defined as follows (for details see \cite{Bondar:2004sv,Braguta:2005kr,Braguta:2006nf}): \\
{\bf $^3S_1$ meson}
\beq
{\langle V_{\lambda}(p)|{\bar Q}_{\beta}(z)\,Q_{\alpha}(-z)|0\rangle}_{\mu}=
\frac{f_{V} M_V}{4}\int^1_o dx_1\,e^{i(pz)(x_1-x_2)}
\biggl \{ {\widehat p}\,\frac{(e_{\lambda}z)}{(pz)}\, V_L(x) + \left ( {\widehat e}_{\lambda} - {\widehat p}\,\frac{(e_{\lambda}z)}{(pz)} \right ) \, V_{\perp}(x)+\nonumber \\
 \frac {2 {\overline M_Q} } {M_V^2} Z_t (\sigma_{\mu\nu} e_{\lambda}^{\mu}\, p^{\nu})\,V_{T}(x)+  
\frac 1 2 \left( 1- Z_t Z_m \frac {4 {\overline M_Q^2}} {M_V^2} \right ) (\epsilon_{\mu\nu\alpha\beta}\gamma_{\mu}\gamma_5\,e_{\lambda}
^{\nu}\,p^{\alpha}z^{\beta})\, V_{A}(x) \biggl \}_{\alpha\beta},
\eeq
{\bf $^1S_0$ meson}
\beq
{\langle P(p)|{\bar Q}_{\beta}(z)\,Q_{\alpha}(-z)|0\rangle}_{\mu}=i
\frac{f_{P} M_P}{4}\int^1_o dx_1\,e^{i(pz)(x_1-x_2)}
\biggl \{  \frac {\hat p \gamma_5} {M_P} P_A(x) - \frac {M_P} {2 {\overline M_Q} } Z_p \gamma_5 P_P(x) 
\biggr \}_{\alpha\beta},
\eeq
{\bf $^3P_0$ meson}
\beq
{\langle \chi_{c0}(p)|{\bar Q_{\beta}}(z) Q_{\alpha} (-z)|0\rangle}_{\mu} = \frac {f_V^{(1)} M_{\chi}} 4 \int d y e^{i pz (x_1 - x_2)} 
\biggl \{ \frac {\hat p} {M_{\chi}} Z_v S_V (x) - 3 \frac {M_{\chi}} {  2 {\overline M_Q}}  Z_p S_S(x)  \biggr \}_{\alpha\beta} .
\eeq
Here the following designations were used: $\overline M_Q = M^{\overline {MS} }_Q ( \mu = M^{\overline {MS} }_Q)$, $x=x_1, x_2=1-x$ are the fractions of meson 
momentum carried by $c$ quark and $\bar c$ antiquark correspondingly, the factors $Z_p, Z_t, Z_m, Z_v$ 
are defined as
\beq
Z_p  &=& \left[\frac{\alpha_s(\mu^2)}{\alpha_s({\overline M}_Q^2)} \right]^{\frac{-3C_F}{b_o}},\quad ~~ 
Z_{t} = \left[\frac{\alpha_s(\mu^2)}{\alpha_s({\overline M}_Q^2)} \right]^{\frac{C_F}{b_o}},\quad 
Z_{m} = \left[\frac{\alpha_s(\mu^2)}{\alpha_s({\overline M}_Q^2)} \right]^{\frac{3C_F}{b_o}}, \quad
Z_v =  \left[\frac{\alpha_s(\mu^2)}{\alpha_s({\overline M}_Q^2)} \right]^{\frac{ 8 C_F}{9 b_o}},
\eeq
where $C_F=4/3$, $b_o=25/3$. The wave functions $\phi_i=V_L, V_{\perp}, V_T, V_A, P_A, P_P, S_S$ are normalized 
as follows: $\int_0^1 d x \phi_i=1$. The wave function $S_V$ is normalized as $\int_0^1 d x (x_1-x_2) S_V(x)=1$.
The dependence of the light cone wave functions on the scale $\mu$ is very slow and 
it will not be considered in the full form in all functions used in our calculation. Only renormalization factors of the corresponding local currents
$Z_p, Z_t, Z_v$ will be regarded. 

Unfortunately today there is no information about the light cone wave function obtained directly from QCD Lagrangian. 
So to proceed with numerical analysis we are forced to use some model for these wave functions.
To find the leading twist wave functions $V_L(x), V_T(x), P_A(x), S_V(x)$ we will apply Brodsky-Huang-Lepage(BHL) \cite{Brodsky:1981jv} procedure 
which allows one to connect the light cone wave functions of leading twist with the equal 
time wave function in the rest frame. 
The equal time wave functions of charmonium mesons will be taken from 
the solution of Schrodinger equation with Buchmuller-Tye potential \cite{Buchmuller:1980su}.
Having these wave functions in momentum space $\psi({\bf k}^2)$ one 
can get the light cone wave functions of leading twist using the following rule \cite{Chernyak:1983ej}:
\beq
\phi_i \sim \int^{{\bf k}_{\perp}^2< \mu^2} d^2 k_{\perp} \psi_c(x,\bf {k}_{\perp}),
\label{p1}
\eeq
where $\psi_c(x,\bf {k}_{\perp})$ can be  obtained from $\psi( {\bf k}^2)$ after the 
substitution \cite{Brodsky:1981jv}
\beq
{\bf k}_{\perp} \to {\bf k}_{\perp}, \quad k_z \to ( x_1 - x_2) \frac {M_0} 2, \quad M_0^2 = \frac {m_c^2 + {\bf k}_{\perp}^2 } {x_1 x_2}.
\eeq
Here $m_c$ is the quark mass in the potential model. 

\newpage
\begin{figure}[t]
\begin{picture}(150, 50)
\put(-40,-420){\epsfxsize=15cm \epsfbox{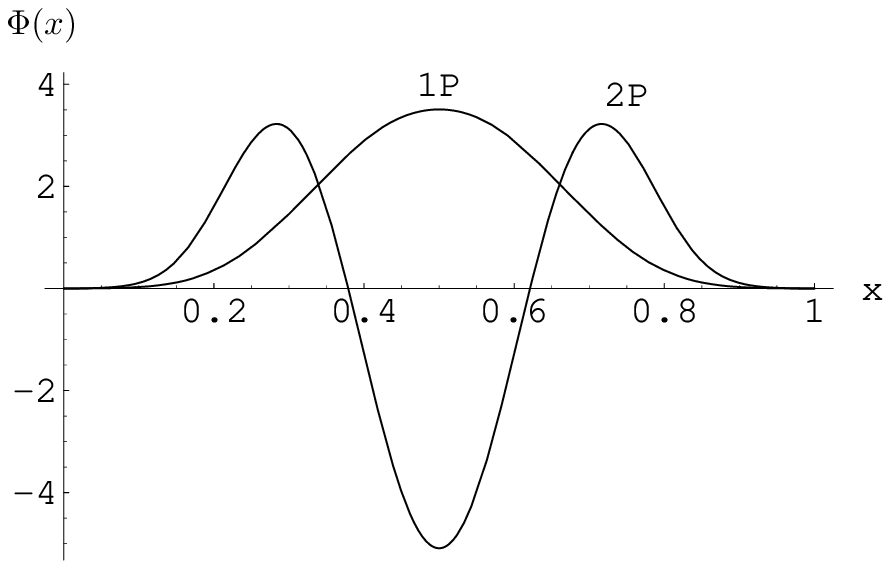}}
\put(-50, -60){\bf{Fig. 2: The functions $\Phi(x)$ for 1P, 2P state mesons.}}
\end{picture}
\end{figure}
\vspace*{0cm}
\vspace*{1.5cm}

It is worth noting that in paper \cite{Bodwin:2006dm} the relations between the light cone wave functions 
and equal time wave functions of charmonium mesons in the rest frame were derived. The procedure proposed in paper \cite{Bodwin:2006dm}
is similar to BHL with the difference: in formula (\ref {p1}) one must make the substitution 
$d^2 k_{\perp} \to d^2 k_{\perp} \sqrt {{\bf k}^2+m_c^2}/(4 m_c x_1 x_2)$. But this substitution was derived at leading 
order approximation in relative velocity of quark-antiquark motion inside the charmonium. At this approximation 
${\bf k}^2 \sim O(v^2),~ 4 x_1 x_2 \sim 1+ O(v^2)$ and the substitution amounts to $d^2 k_{\perp} \to d^2 k_{\perp} (1 + O( v^2 ))$. 
Thus at leading order approximation applied in \cite{Bodwin:2006dm} these two approaches coincide. 

Using equation (\ref{p1}) the expression for the leading twist wave 
functions can be easily written in the following form
\beq
\phi_i(x) = c_i \phi^{as}_i (x) \Phi_i (x),
\label{func}
\eeq
where the constants $c_i$ are fixed from the normalization condition, $\phi^{as}_i (x)$ are the 
asymptotic forms of wave functions of leading twist\footnote {for the wave functions $V_L, V_{\perp}, V_T, V_A, P_A, P_P, S_S$ $\phi^{as} \sim x_1 x_2$, 
for the wave function $S_V$ $\phi^{as} \sim x_1 x_2 (x_1-x_2)$. The asymptotic forms of the other functions used in the calculation can be found in papers
 \cite{Bondar:2004sv,Braguta:2005kr,Braguta:2006nf}}, the functions $\Phi_i(x)$ are given by the formulas
\beq
\nonumber
\Phi(x) &=& \int_0^{\frac {\mu^2} {4 x_1 x_2}} d \xi \psi( \xi + \frac {(x_1 -x_2)^2} {4 x_1 x_2} m_c^2 ),
\eeq
for S-wave mesons, 
\beq
\Phi(x) &=& \int_0^{\frac {\mu^2} {4 x_1 x_2}} d \xi \sqrt {\frac { \xi + \frac {m_c^2} {4 x_1 x_2} } { \xi + \frac {m_c^2} {4 x_1 x_2}(x_1-x_2)^2 }   }  \psi( \xi + \frac {(x_1 -x_2)^2} {4 x_1 x_2} m_c^2 ),
\label{Phi}
\eeq
for P-wave mesons. 

For the light cone wave functions of nonleading twist there is no relation similar to equation (\ref{p1}) and 
to calculate these functions the following model will be applied. It is known that nonleading twist wave functions can be written as a product \cite{Chernyak:1983ej}
\beq
\phi_i(x) \sim \phi^{as}_i (x) {\tilde \Phi}_i(x), 
\eeq
with unknown functions ${\tilde \Phi}_i(x)$. In our calculation we will suppose that the functions ${\tilde \Phi}_i(x)$ equal to the corresponding 
functions $\Phi_i(x)$ (\ref{func}) of leading twist. Thus the model for the light cone wave functions of leading and nonleading twist 
is given by equations (\ref{func})-(\ref{Phi}).

\section{The study of the process $e^+e^- \to J/ \Psi X(3940)$.}
First let us consider the  hypothesis: {\bf $\bf X(3940)$ is one of $\bf \chi'_{c0}(2^3P_0), \chi'_{c1}(2^3P_1), \chi'_{c2}(2^3P_2)$ mesons}. 
At Belle $X(3940)$ is seen to decay to $D {\overline D^*}$\cite{Abe:2005hd} and not to $D {\overline D}$. 
If this $D {\overline D^*}$  is dominant decay mode than $X(3940)$ is $\chi'_{c1}(2^3P_1)$. Unfortunately the 
process $e^+ e^- \to J/ \psi \chi'_{c1}(2^3P_1)$ has not been  considered in the framework of 
light cone formalism yet but this formalism tells us that the cross section of the process $e^+ e^- \to J/ \psi \chi'_{c1}(2^3P_1)$ is 
suppressed by the factor $1/s$ in comparison with the cross section of the process $e^+ e^- \to J/ \psi \chi'_{c0}(2^3P_0)$ \cite{Braaten:2002fi}. 
So one can expect that the cross section of $\chi'_{c0}$ production is greater than the cross section of $\chi'_{c1}$ production 
and if  $\chi'_{c1}$ is seen at the experiment $\chi'_{c0}$ meson must be 
seen also. But the decay mode of $\chi'_{c0}$ meson is $D \bar D$ and since this decay mode is not seen at the experiment 
we reject this hypothesis. 

\newpage
\begin{figure}[t]
\begin{picture}(150, 50)
\put(-120,-70){\epsfxsize=6cm \epsfbox{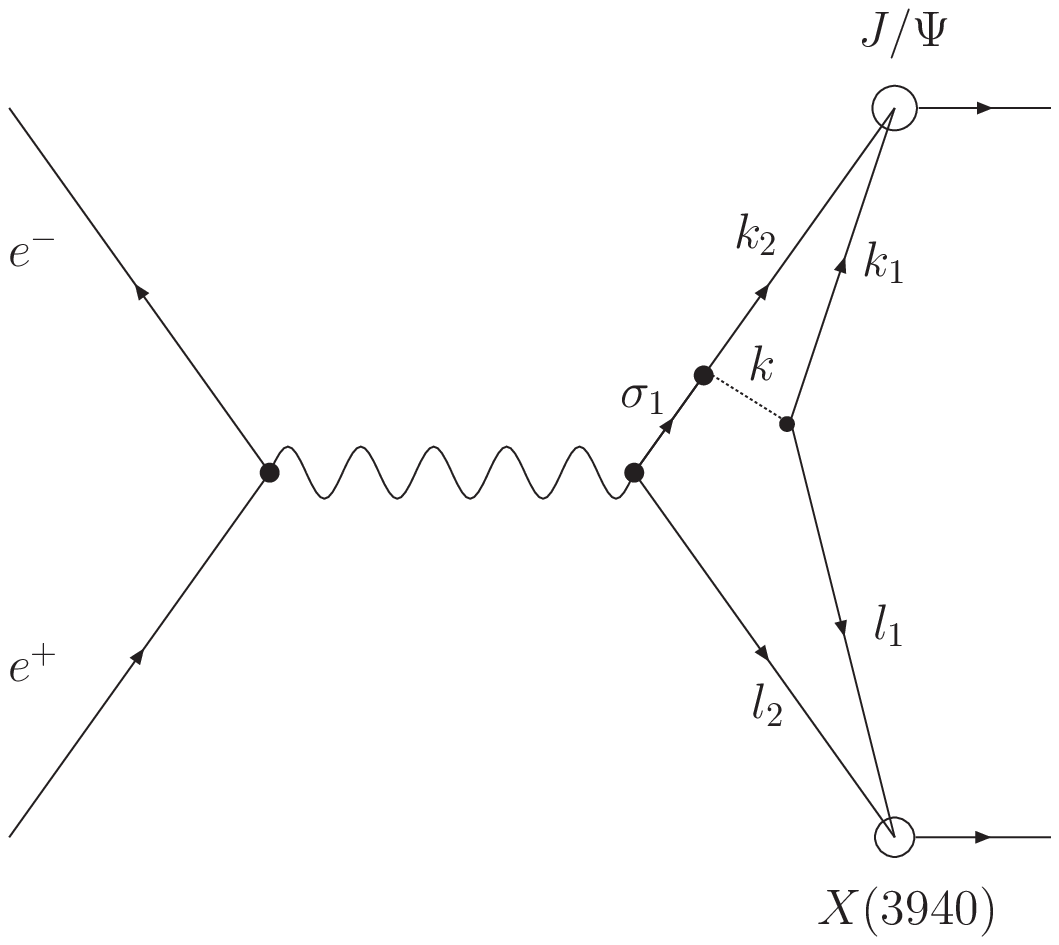}}
\put(100,-70){\epsfxsize=6cm \epsfbox{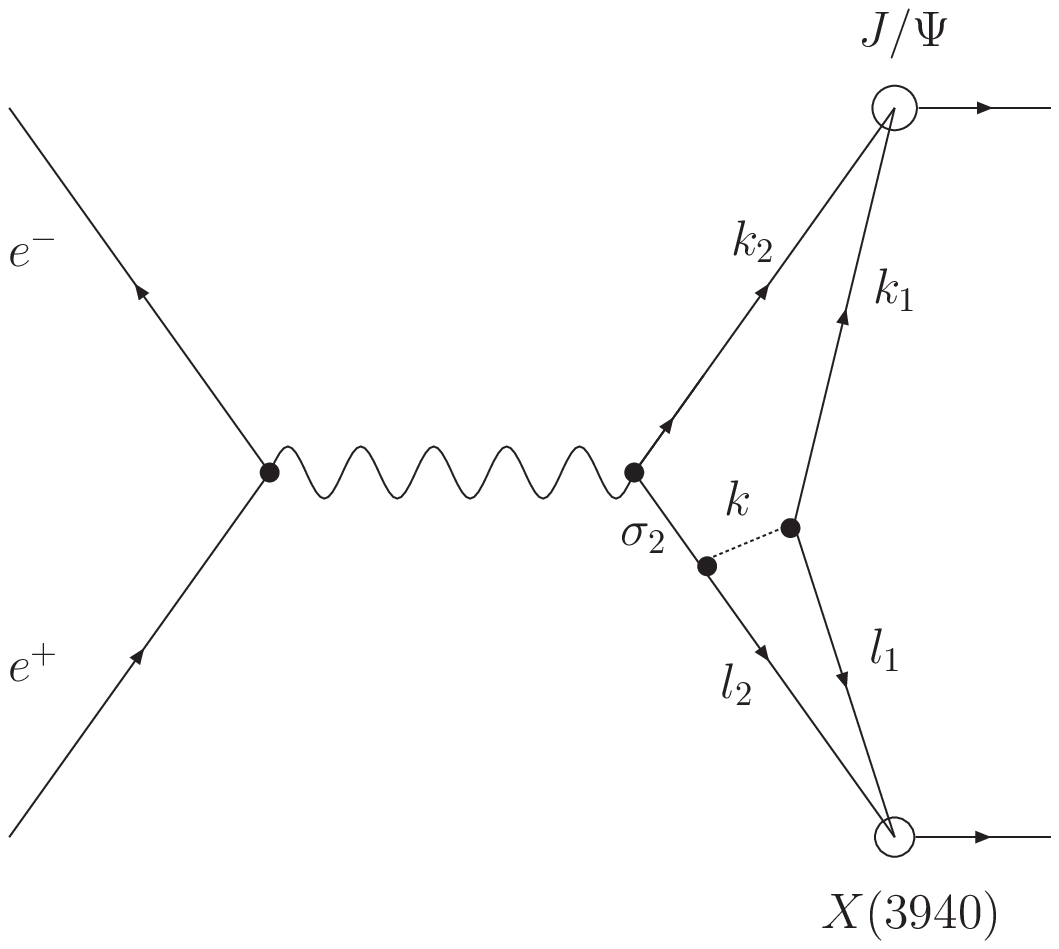}}
\put(-100, -85){\bf{Fig. 3: The diagrams that contribute to the process $\epem\to J/ \psi X(3940)$.}}
\end{picture}
\end{figure}
\vspace*{2.5cm}

Contrary to the NRQCD predictions the values of the cross sections of exited charmonium mesons production measured at Belle and BaBar 
experiments are rather large. For instance due to this the process $\epem \to J/\Psi \eta_c'$ with excited 
$\eta_c'$ meson production is seen at the experiments. In connection with this fact the question arises: 
if the process $\epem \to J/\Psi \chi_{c0}$ is seen at the experiment and $X(3940)$ is not $\chi'_{c0}$ meson 
why one does not see the production of excited charmonium state $\chi'_{c0}$. One of the possible answers to this question is 
presented in Fig. 2 where the functions $\Phi(x)$ (\ref{Phi}) for 1P and 2P states are 
shown. From this plot one can see that contrary to 1P state the function of $2P$ state $\Phi(x)$ is oscillating function.   
In light cone formalism this wave function must be integrated with the hard part of the amplitude of the process, 
consequently one can expect considerable cancellation in the amplitude.
We will not estimate the value of the cross section of $\chi'_{c0}$ meson production since the accuracy of our 
calculation for this process is very low. This can be understood as follows if 
we divide the amplitude of $\chi'_{c0}$ meson production into two parts where functions $\Phi_{2P}$ 
is positive and where this functions is negative than the amplitude is the difference between two 
large close numbers. Evidently the error of such calculation is large. It should be noted here that 
if our conjecture is correct and the suppression of $\chi'_{c0}$ production really results from the oscillation in the wave function 
of this meson than one can expect that the production of $\chi'_{c0}, \chi'_{c1}, \chi'_{c2}$ mesons 
is suppressed in any hard process. 

Now let us consider the hypothesis: {\bf $\bf X(3940)$ is excited $0^{-+}$ meson}. 
Two diagrams that give contribution to the amplitude of the process under consideration are 
presented in Fig 3. The other two can be obtained from the depicted ones by charge conjugation. 
The amplitude of the process $e^+e^- \to J/ \psi P$ for final mesons with equal masses was first found in paper \cite{Bondar:2004sv}.
In our calculation the result of paper \cite{Braguta:2005kr} will be used where the case of different masses was considered. 
The cross section of the process involved can be expressed through the formfactor $F_{vp}$ defined as follows
\beq
\left<V(p_1, \lambda), P(p_2)| J_{\mu} | 0\right> & = & 
  \epsilon_{\mu \nu \rho \sigma} e^{\nu} p_1^{\rho} p_2^{\sigma} F_{vp},
\label{Fvp}
\eeq
The formfactor $F_{vp}$ equals 
\beq
|F_{vp}(s)| & = & \frac{32\pi}{9}   \left|\frac{f_V f_P  M_P M_V}{q_0^4}\right|\,I_0\,,
\label{res}
\eeq
\beq
I_0 & = &
  \int^1_0 dx_1 \int^1_0 dy_1 \alpha_s(\mu^2) \left\{
    \frac {M_P} {M_V^2} \frac{Z_t Z_p 
    V_{T}(x) P_{P}(y)}{d(x,y)\, s(x)}- \frac 1 {M_P} \frac{\overline {M}_Q^2 }{{ M_V}^2}\,
    \frac{Z_m(\mu^2) Z_t V_T(x) P_A(y)}{d(x,y)\,s(x)}+
\right. \nonumber \\ & + &
\frac{1}{2 M_P}\frac{V_{L}(x)\,P_{A}(y)}{d(x,y)}+\frac{1}{2 M_P}
\frac{(1-2y_1)}{s(y)}\frac{V_{\perp}(x)\,P_{A}(y)}{d(x,y)}+
\nonumber \\ & + & \left.
\frac{1}{8} \biggl ( 1-Z_t Z_m \frac{4{ \overline M}_Q^2}{{ M_V}^2 }\biggr ) \frac 1 {M_P} \, 
\frac{(1+y_1)V_A(x)P_A(y)}
{d^2(x,y)}\right\},
\label{I0}
\eeq
where $q_0^2 \simeq (s-M_V^2-M_P^2)$, $P_A, P_P, V_T, V_L, V_{\perp}, V_A$ are 
the light cone wave functions defined above, $M_V, M_P$ are the mass of the 
vector and pseudoscalar mesons correspondingly, $d(x,y), s(x), s(y)$ are defined as follows:
\beq
d(x,y)& = & 
  \frac{k^2}{q_0^2}=\left( x_1+\frac{\delta}{y_1}\right)
  \left(y_1+\frac{\delta}{x_1}\right),
\qquad \delta=\frac { ( Z_m  {{\overline  M}_Q} )^2} {q_0}\,, 
\label{gluon}
\eeq
%\vspace*{1.5cm}
\beq
s(x) & = & \left(x_1+\frac{(Z_m {\overline  M}_Q)^2}{y_1y_2\,q_0^2} \right),
\quad s(y)= \left(y_1+\frac{(Z_m {\overline  M}_Q)^2}{x_1x_2\,q_0^2}
\right).
\label{quark}
\eeq

\vspace*{1.5cm}

From formula (\ref{gluon}) one sees that gluon propagator $d(x,y)$ tends to infinity when the momenta
of quark (${\bf k_1}$) or antiquark (${\bf l_1}$) created by this gluon tend to zero. Evidently this property has nothing to 
do with real situation: $d(x,y) \to 4 m_c^2/q_0^2$ when $x,y \to 0$. Similar problem is seen in 
the expressions for quark propagators $s(x), s(y)$. This problem appeared since the authors of 
paper \cite{Bondar:2004sv} used the following expression for the quark momenta $k_1 =(+, \perp, -) = ( q_0 x_1, 0, M_Q^2/x_1 q_0)$ ( and 
similarly for three other quarks). It is not difficult to understand that this expression can be applied if 
the energy of the quark is $\sim \sqrt s$. If $x \to 0$ the energy of the quark is $\sim q_0 x_1 \ll \sqrt s$ and this 
expression becomes inapplicable. Nevertheless the calculation of the cross section of $1S, 2S$ charmonium mesons 
production can be done using equations (\ref{Fvp})-(\ref{quark}) since these charmonium mesons can be considered 
as nonrelativistic objects. This means that the regions where expressions (\ref{gluon}), (\ref{quark}) become 
incorrect are suppressed. 

Contrary to the $1S$ and $2S$ charmonium mesons there is considerable contribution to the cross section $3S$ charmonium meson 
production from the regions where the expressions (\ref{gluon}), (\ref{quark} are incorrect. Thus these expressions must be modified. 
We will do this as follows. First we will disregard transverse motion of quark-antiquark pair even in the end point region $x,y \sim 0$ 
since account of the transverse motion is higher order $v$ effect. Next if the energies of all quark $q_0 x_i, q_0 y_i$ in Fig.3 are greater 
than $m_c$ than the following expression for the quarks and gluon propagators will be used:
\beq
\nonumber
d(x,y) &=& x_1 y_1, \\ \nonumber
s(x) &=& x_1, \\ \nonumber
s(y) &=& y_1.
\eeq
If the energy $q_0 x_1$ of quark $k_1$ reaches the value $m_c$, what means that this quark is at rest, 
the expressions of the quark momentum  $k_1 =( q_0 x_1, 0, m_c^2/x_1 q_0)$ must be substituted by $(m_c, 0, m_c)$
(and similarly for three other quarks). We will not write explicit expressions for gluon and quark 
propagators $d(x,y), s(x), s(y)$ since they are rather cumbersome. The model that will be applied in 
our calculation is rude but it is sufficient to apply it for the estimation of the value of the 
cross section. It is worth to note  here that in the calculation of quark propagators $s(x), s(y)$ we 
distinguish running mass $M_Q(\mu)$ inside the quark propagators from the pole mass $m_c$ of the external 
quarks $k_1, k_2, l_1, l_2$. Moreover in our calculation all scale dependent quantities will be taken at scale $\mu^2 \sim s/4$.

\begin{figure}[t]
\begin{picture}(150, 50)
\put(-40,-420){\epsfxsize=15cm \epsfbox{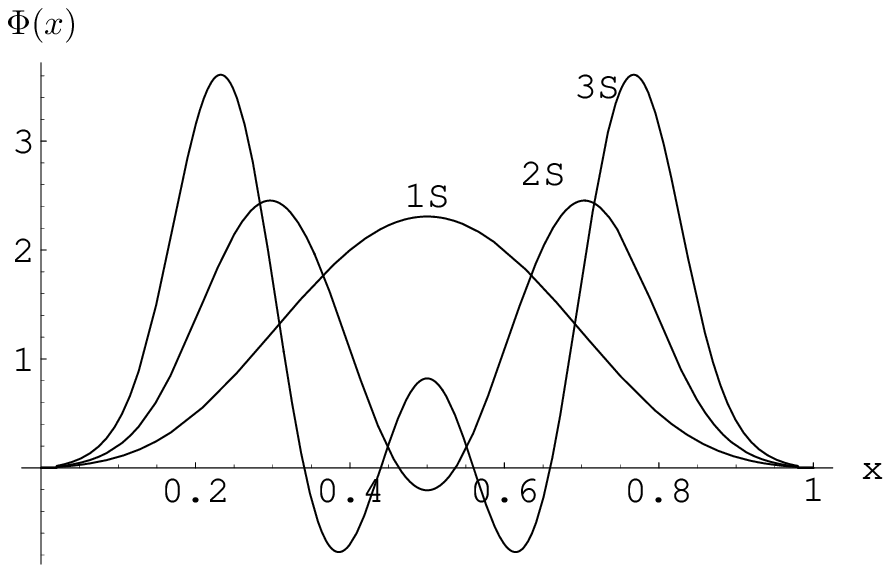}}
\put(-50, -60){\bf{Fig. 4: The functions $\Phi(x)$ for 1S, 2S, 3S state mesons.}}
\end{picture}
\end{figure}

There are two contributions to the  formfactor $F_{vp}$ factored in formula (\ref{res}).
The first contribution originates from the wave function of mesons at the origin and it is 
proportional to $\sim f_V f_P$.  The second contribution regards internal motion 
of quark-antiquark pair inside mesons and it is proportional to $I_0$. The leading
NRQCD approximation does not take into the account the contribution of 
the second type. 

In the numerical calculation one loop expression for $\alpha_s$ with $\Lambda_{QCD}=200$MeV will be used, ${\overline M}_Q=1.2$GeV, $m_c=1.4 \pm 0.2$. 
The functions (\ref{Phi}) of 1S, 2S, 3S meson states at $\mu \sim {\overline M}_Q$  are presented in Fig. 4.
The constant $f_V=0.4$GeV is determined from the decay width $\Gamma(J/ \psi \to e^+ e^-)$. Supposing
that $X(3940)$ is $3^1S_0$ meson the constant $f_P$ can be estimated as $f_P \simeq f_V(3^3S_1)$. In turn
the constant $f_V(3^3S_1)$ can be found from the decay width $\Gamma(\psi(4040) \to e^+ e^-)$. 
Thus we have $f_P \simeq 0.17$GeV. These numerical parameters lead to the value of the cross section 
\beq
\sigma(e^+e^- \to J/ \psi X(3940)) \simeq  11 \pm 3 ~ \mbox{fb}.
\label{sig1}
\eeq
Experimental result for this cross section is $\sigma \times Br(X(3940)>2 \mbox{charged particles}) = (10.6 \pm 2.5 \pm 2.4)$fb\cite{Abe:2005hd}
and it is in good agreement with our prediction. Sure there are a number of uncertainties connected with 
different reasons. For instance, one does not know the size of QCD radiative corrections and the size of $1/s$ corrections. 
In addition to these uncertainties there is an error due to the model of the light cone wave functions (\ref{func}) used 
in our paper. 

To calculate equal time charmonium wave functions nonrelativistic potential model was applied.
Our calculation shows that there is considerable contribution to the cross sections from the kinematical region 
where the motion of quark-antiquark pair inside charmonium cannot be regarded as nonrelativistic. 
To make our prediction more reliable in addition to nonrelativistic potential model we 
have calculated the cross section in the framework of the model 
where charmonium mesons are treated as a relativistic quark-antiquark bound state \cite{Anisovich:2005vs}.
In the framework of this model the value of the cross section is $\sigma \sim 10 \pm 3$ fb.

\section{Discussion.}

In our paper light cone formalism was applied to the process $\epem \to J/\Psi X(3940)$. 
We considered two hypotheses about the structure of $X(3940)$ meson: $X(3940)$
is one of $\chi'_{c0}, \chi'_{c1}, \chi'_{c2}$ mesons and 
$X(3940)$ is $\eta''_c$ meson. Based on experimental data we rejected the first 
hypothesis and proposed the conjecture why  $\chi'_{c}$ mesons 
are not seen at the experiment. If this conjecture is correct we predict the suppression 
of $\chi'_{c}$ mesons production in any hard process. As to the second hypothesis 
we calculated the cross section of $X(3940)$ meson production if this meson is $\eta''_c$. 
The result of our calculation is in good agreement with the experiment. 

There are a number of uncertainties of our calculation connected with different reasons. For instance, 
one does not know the size of QCD radiative corrections and the size of $1/s$ corrections. 
But we believe that the main source of uncertainty is our model for the light cone wave functions. 
The main problem is that in our calculation we used nonrelativistic potential model. 
At the same time there is considerable contribution to the amplitude under consideration 
from region where the motion of quark-antiquark pair inside  mesons cannot be considered 
as a nonrelativistic. Obviously application of nonrelativistic potential models to this 
region results in large error. Nevertheless we believe that if one was able to build 
correct description of chamonium in the relativistic region the value of the cross section (\ref{sig1})
would be of order of $10$ fb. So the estimation  of the cross section (\ref{sig1}) using nonrelativistic 
potential model is rather good. 

Last statement is based on the following facts. First of all in addition to the nonrelativistic 
potential model we applied relativistic potential model \cite{Anisovich:2005vs} to the calculation of the cross 
section of the process $e^+e^- \to J/ \psi X(3940)$. In the framework of this 
model we obtained the value $\sigma \sim 10 \pm 3$ fb. Moreover it is possible to 
estimate the cross section in a model independent way. As was noted above the amplitude of the 
process $\epem \to J/ \Psi P$ is a product of NRQCD result for this amplitude and the 
factor that regards internal motion of quark antiquark pair inside the mesons. For 
the cross section this statement can be written in the following form
\beq
\sigma(\epem \to J/ \Psi P) \sim f_P^2 |I_0(P)|^2.
\label{est}
\eeq
Obviously the wave functions of higher charmonium states become wider. Consequently the factor $I_0(P)$
that regards internal motion of quark antiquark pair inside the mesons is larger for higher 
charmonium states. So $I_0(3S) > I_0(2S)$. Using this relation and equation (\ref{est}) 
one can get lower bound for the cross section of $\eta''_{c}$ meson production
\beq
\sigma( \epem \to J/ \Psi ~ \eta''_{c} ) > \sigma( \epem \to J/ \Psi ~ \eta'_{c}) \frac {f_{\eta''_c}^2} { f_{\eta'_c}^2 } \sim 6 ~\mbox{fb}.
\eeq
Another estimation of the cross section can be obtained  if one supposes that $I_0(2S)/I_0(1S) \sim I_0(3S)/I_0(2S)$.
This estimation gives $\sigma \sim 8$ fb.

As it was noted earlier the motion of quark antiquark pair near to the end point regions($x \sim 0, x \sim 1$) 
is relativistic and cannot be calculated reliably in the framework of nonrelativistic
potential models. Recently in paper \cite{Bodwin:2006dm} it was proposed to solve this 
problem in the framework of NRQCD. We are not going to discuss this paper in 
detail. We would like to say only that the results of \cite{Bodwin:2006dm} were not 
applied in our paper since we don't think that it is possible to regard 
relativistic motion as it was done in \cite{Bodwin:2006dm}.

The interpretation of $X(3940)$ as $3^1S_0$ state was proposed in paper \cite{Rosner:2005gf}. 
The problem with this interpretation consists in the fact that the mass of this meson 
obtained in the framework of potential approach is $4040-4060$MeV\cite{Swanson:2006st}.
But as it was shown in our paper relativistic motion of quark antiquark pair is important 
for $3^1S_0$ meson and probably potential models do not regard this motion correctly.
Moreover in paper \cite{Gershtein:2006ng} it was shown that $X$ meson with the mass $3940$MeV agrees well 
with Regge trajectory with quantum numbers $0^{-+}$. 

This work was partially
supported by Russian Foundation of Basic Research under grant 04-02-17530, Russian Education
Ministry grant E02-31-96, CRDF grant MO-011-0, Scientific School grant SS-1303.2003.2. One of
the authors (V.B.) was also supported by Dynasty foundation.

\end{document}